\documentclass[aps,pra,reprint,superscriptaddress,letterpaper,showpacs]{revtex4-1}
\usepackage[english]{babel}
\usepackage{ucs}
\usepackage[utf8x]{inputenc}
\usepackage{amsmath}
\usepackage{amsfonts}
\usepackage{amssymb}
\usepackage{graphicx}
\usepackage{bm}
\usepackage{bbm}
\usepackage{braket}
\usepackage{color}

\begin{document}

\title{Dispersive readout of valley splittings in cavity-coupled silicon quantum dots}

\author{Guido Burkard}
\affiliation{Department of Physics, University of Konstanz, D-78457 Konstanz, Germany}
\author{J. R. Petta}
\affiliation{Department of Physics, Princeton University, Princeton, New Jersey 08544, USA}

\pacs{03.67.Lx, 73.63.Kv, 85.35.Gv}

\begin{abstract}
The bandstructure of bulk silicon has a six-fold valley degeneracy. Strain in the Si/SiGe quantum well system partially lifts the valley degeneracy, but the materials factors that set the splitting of the two lowest lying valleys are still under intense investigation. We propose a method for accurately determining the valley splitting in Si/SiGe double quantum dots embedded into a superconducting microwave resonator. We show that low lying valley states in the double quantum dot energy level spectrum lead to readily observable features in the cavity transmission. These features generate a ``fingerprint" of the microscopic energy level structure of a semiconductor double quantum dot, providing useful information on valley splittings and intervalley coupling rates.


\end{abstract}

\maketitle


\section{Introduction}

Silicon is a promising material system for spin-based quantum information processing due to weak spin-orbit and hyperfine couplings \cite{Zwanenberg2013}. Electron spin lifetimes as long as $T_1$ = 3000 seconds were measured as early as 1959 in phosphorous doped silicon \cite{FeherT11959}. In natural abundance silicon, electron spin coherence times $T_2$ = 60 ms have been reported \cite{TyryshkinPRB2003}. Isotopic enrichment has extended the quantum coherence time to $T_2$ = 10 seconds \cite{TyryshkinNatMat2012}. Moreover, the ability to dope silicon with a wide range of donors and acceptors is particularly exciting, as heavy elements such as $^{209}$Bi (with nuclear spin quantum number $I$ = 9/2) have a complicated energy level structure that results in so-called ``clock transitions" that are first-order-insensitive to magnetic field fluctuations \cite{Wolfowicz2013}. Coupling to the nuclear spin of a single phosphorous donor also allows access to an additional quantum degree of freedom that can be used as a long-lived quantum memory \cite{Pla2013,Saeedi2013}.

In terms of its ability to support quantum coherence, the trajectory of the silicon material system is quite impressive \cite{Awschalom2013}. On the other hand, silicon presents severe materials challenges in quantum devices, where control at the level of single electrons is desired. Electrons confined in Si/SiGe quantum wells have an effective mass $m^*$ = 0.19 $m_e$ (roughly three times larger than the GaAs/AlGaAs quantum well system), where  $m_e$ is the free electron mass \cite{Ando1982,Schaffler1997}. As a result, Si quantum devices must be significantly smaller than their GaAs counterparts to achieve similar orbital excited state energies. Over the past several years, the effective mass challenge has been effectively solved through the development of novel overlapping gate architectures, and the isolation of single electrons in accumulation mode Si/SiGe quantum dots (QD) is becoming routine \cite{Angus2007,Borselli2015,Veldhorst2015,Zajac2015,Zajac2016}.

A major remaining challenge is to understand the factors that limit the valley splitting in silicon \cite{Friesen2010}. The bulk electronic bandstructure of Si has six equivalent minima (termed valleys) that are located 0.85 of the way from the Brillouin zone edge \cite{Ando1982}. In Si/SiGe quantum well systems, the 4\% larger lattice constant of Ge strains the Si quantum well, raising in energy the four in-plane $\Delta_4$ valleys and lowering in energy the two perpendicular-to-the-plane $\Delta_2$ valleys \cite{Schaffler1997}.  In view of the interplay between the spin and valley degrees of freedom \cite{Rohling2012}, the ability to probe and, ultimately, control the splitting between the remaining quasi-degenerate valleys in Si/SiGe 
quantum well systems represents an urgent challenge on the way towards scalable spin qubits in Si/SiGe QDs.

Theory suggests that the vertical electric field sets the overall scale of the valley splitting \cite{Uemura1977}; a prediction that has been experimentally verified in Si MOS (metal-oxide-semiconductor) QDs \cite{Yang2013}. However, in the Si/SiGe system, the valley splitting is known to substantially vary in QD devices fabricated on the same heterostructure. Work by Borselli \textit{et al.}\ reports valley splittings in the range of 120 to 270 $\mu$eV \cite{Borselli2011}. In recent work by Zajac \textit{et al.}, valley splittings in the range of 35 -- 70 $\mu$eV were extracted in the same multiple QD device \cite{Zajac2015}. Measurements by Shi \textit{et al.}\ show that the valley splitting can be tuned by using gate voltages to laterally shift the position of the electronic wave function in the two-dimensional electron gas \cite{Shi2011}. These experiments suggest that the microscopic structure of the QD system (interface roughness, step edges, etc.) plays a strong role in determining the valley splitting \cite{Friesen2006}. Unfortunately magnetospectroscopy measurements are time consuming to perform and the data can often be ambiguous, especially when the valley splitting is of the order of $k_{\rm B}T$,  where $k_{\rm B}$ is Boltzmann's constant and $T$ is the electron temperature. Therefore the development of new probes of valley splitting will benefit the QD community. 

In this paper we propose a cavity-based measurement of the low lying energy level structure of few-electron semiconductor double quantum dots (DQD) in the circuit quantum electrodynamics (cQED) architecture. Hybrid DQD-cQED systems have been used to demonstrate electric dipole couplings $g_0/2\pi$ ranging from 10 to 100 MHz \cite{Delbecq2011,Frey2012,Petersson2012,Deng2015}, quantum control and readout of spin-orbit qubits \cite{Petersson2012}, and spin-photon coupling \cite{Viennot2015}. In essence, these experiments probe the electric susceptibility $\chi$ of a mesoscopic system with a sensitivity well-beyond that of a single electron \cite{Stehlik2015}. The susceptibility is the largest at a DQD interdot charge transition, where a single electron can tunnel from the left dot to the right dot, resulting in an electric dipole moment that is roughly 1000 times larger than in atomic systems \cite{Frey2012}. Here we show that the low lying valley structure of a few electron Si/SiGe QD is directly accessible in a hybrid cQED system. The amplitude response of the cavity generates a fingerprint of the DQD energy level structure, providing not only access to the energy level splittings, but also the interdot and intervalley coupling rates. We model the system response using realistic parameters that should be accessible in future experiments.


\section{Model}

\subsection{Cavity-coupled double quantum dot}

\begin{figure}
\begin{center}
\includegraphics[width=\columnwidth]{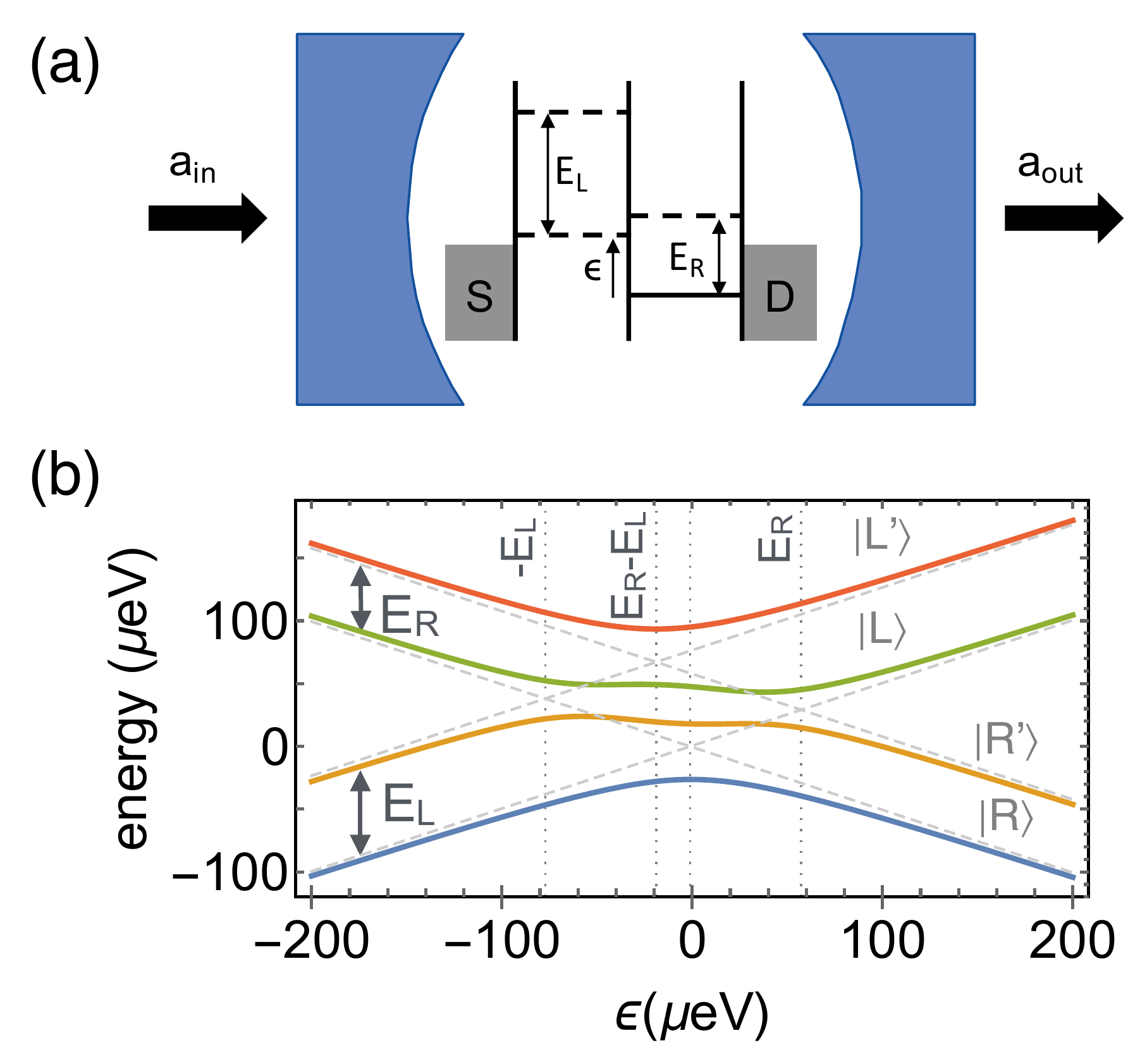}
\caption{(a) A cavity-coupled DQD is probed using a small input field $a_{\rm in}$ with frequency $\omega_{\rm R}$. Charge dynamics in the DQD result in changes in the transmitted field $a_{\rm out}$. The DQD is coupled to source (S) and drain (D) electrodes.  (b) DQD energy levels plotted as a function of energy level detuning $\epsilon$. In general, the left dot valley splitting $E_{L}$ is different than the right dot valley splitting $E_{R}$.  For this plot, $E_L$ = 76 $\mu$eV, $E_R$ = 58 $\mu$eV, $t = 25\,\mu {\rm eV}$, and $t' = 13\,\mu {\rm eV}$.}
\label{fig:cartoon}
\end{center}
\end{figure}

Figure~\ref{fig:cartoon}(a) illustrates the proposed experimental system. A DQD containing a single excess electron is electric-dipole coupled to a high quality factor superconducting resonator. We model the DQD as a four-level system consisting of the left dot ground state $|L\rangle  = |(1,0)\rangle$, left dot excited state $|L'\rangle = |(1',0)\rangle$, right dot ground state $|R\rangle = |(0,1)\rangle$, and right dot excited state $|R'\rangle = |(0,1')\rangle$. The left dot valley splitting $E_{L}$ = $E_{|L'\rangle} - E_{|L\rangle}$ is often different than the right dot valley splitting $E_{R}$. The energy difference between the left dot ground state $|L\rangle$ and the right dot ground state $|R\rangle$ is set by the detuning $\epsilon$. In general, this system could be used to measure a variety of low lying excited states, such as orbital excited states, valley states, and Zeeman split states \cite{Viennot2015}. We focus on the Si/SiGe QD system, where valley splittings are typically $<$200 $\mu$eV in energy \cite{Borselli2011,Zajac2015}.

The cavity field is sensitive to charge dynamics in the DQD due to the large electric dipole coupling that is achieved in cQED systems \cite{Frey2012}. In typical experiments, the cavity is probed by driving it with an input field $a_{\rm in}$ with frequency $\omega_{R}$ and detecting the transmitted field $a_{\rm out}$. Measurements of the cavity response provide useful information about the mesoscopic systems (e.g.\ a quantum dot) embedded in the cavity \cite{Delbecq2011,Simon2016}. Both the amplitude and phase of the transmitted signal provide useful information. As an example, sequential tunneling through a voltage biased DQD was recently shown to result in microwave frequency amplification, such that $|a_{\rm out}/a_{\rm in}|$ $>$ 1 \cite{Liu2014,Liu2015,Stockklauser2015}.

\subsection{Hamiltonian}

We model the single electron Si/SiGe DQD using the Hamiltonian,
\begin{align}
& \tilde H_0 =
\begin{pmatrix}
\epsilon/2 + E_{L}	& 0	& t & t'		\\
0	& \epsilon/2 	& -t'  &  t	 \\
t	& -t'	& -\epsilon/2 + E_{R}&0 \\
t'	&t	&0	&-\epsilon/2 \\
\end{pmatrix},
\label{Hdqd}
\end{align}
which is expressed in the local valley eigenbasis $|L'\rangle$, $|L\rangle$, $|R'\rangle$, $|R\rangle$. 
The valley eigenstates $|L\rangle$ and $|L'\rangle$ ($|R\rangle$ and $|R'\rangle$) of the left (right) QD are split by the 
valley splitting $E_L$ ($E_R$). In general, the excited states may have a different projection onto the $\pm z$ valley basis states (see Appendix A).
Therefore, the matrix elements that couple the four levels are distinct.  The states $|L\rangle$ and $|R\rangle$ 
($|L'\rangle$ and $|R'\rangle$) are hybridized by the (intravalley) interdot tunnel coupling $t$ 
near $\epsilon = 0$ ($\epsilon=E_R-E_L$). The valley state $|L'\rangle$ ($|R'\rangle$) is coupled to $|R\rangle$ ($|L\rangle$) by the intervalley matrix element $t'$ leading to avoided crossings near $-E_L$ ($E_R$).

The DQD energy levels are plotted as a function of the detuning parameter $\epsilon$ in Fig.~\ref{fig:cartoon}(b). The left (right) dot energy levels increase (decrease) in energy with increasing $\epsilon$. In Fig.~\ref{fig:cartoon}(b) we take $E_{L}$ = 76 $\mu$eV, $E_{R}$ = 58 $\mu$eV, 
$t= 25\, \mu {\rm eV}$, and $t' =  13\, \mu {\rm eV}$.
 For reference, the cavity frequency $f_0 = 7.8\, {\rm GHz}$ corresponds to an energy of $32 \,\mu {\rm eV}$.

\subsection{Electric dipole coupling}

We assume that the DQD is irradiated with a classical probe field with angular frequency $\omega_R$. The probe field generates an oscillating voltage inside the superconducting resonator. This voltage is directly coupled to the DQD detuning parameter, making it time dependent: $\epsilon(t)$ = $\epsilon_0$ + $\delta\epsilon\cos(\omega_R t)$. Here $\epsilon_0$ is a static energy level detuning, which can be slowly varied in experiments using dc gate voltages. The parameter $\delta\epsilon$ describes the magnitude of the detuning modulation. The interaction with the probe field thus gives rise to a term in the Hamiltonian,

\begin{equation}
\tilde H_P = \frac{1}{2}\delta\epsilon \cos(\omega_R t)\sigma_z,
\end{equation}
where, in the same basis used in Eq.~(\ref{Hdqd}) above,
\begin{equation}
\sigma_z   
= \begin{pmatrix}
1 & 0	& 0 & 0	\\
0	& 1 	& 0	& 0 \\
0	& 0	& -1 &0 \\
0	& 0	&0	&- 1\\
\end{pmatrix}.
\label{Hdipole}
\end{equation}

The coupling of the DQD to a single quantized mode of a microwave cavity with resonance frequency $f_0=\omega_0/2\pi$ can be described as
\begin{equation}
\tilde{H}_I = 2g_0 \left( a+a^\dagger \right) \sigma_z ,
\end{equation}
where $a^\dagger$ and $a$ are the bosonic creation and annihilation operators for the cavity photons
and $a+a^\dagger$ is proportional to the electric field of the cavity mode.  The cavity mode evolves 
according to the Hamiltoninan $\tilde{H}_C = \omega_0 a^\dagger a$ in units where $\hbar=1$. We take $g_0/2\pi$ = 30 MHz in what follows.

In order to describe the dissipative dynamics of the DQD-cavity system, including its steady state,
it is convenient to work in the eigenbasis of $\tilde{H}_0$.  Writing $U_0$ for the unitary operator that
diagonalizes $\tilde{H}_0$, we have
\begin{equation}
  \bar{H}_0 = U_0 \tilde{H}_0 U_0^\dagger = \sum_{n=0}^3 E_n \sigma_{nn} ,
\end{equation}
where $E_0 \le E_1 \le E_2 \le E_3$ are the ordered eigenvalues of $\tilde{H}_0$,
$\sigma_{mn}=|m\rangle\langle n|$, and $|n\rangle$ denotes an eigenstate of $\tilde{H}_0$
with eigenvalue $E_n$.
The dipole operator $\sigma_z$ then needs to be transformed into the eigenbasis
of $\tilde H_0$,
\begin{equation}
D = U_0 \sigma_z U_0^\dagger = \sum_{m,n=0}^3 d_{mn} \sigma_{mn},
\end{equation}
where the matrix elements $d_{mn} = d_{nm}^*$ determine the
dipole transition matrix elements between energy eigenstates.
When transforming the full Hamiltonian $\tilde{H} = \tilde{H}_0 + \tilde{H}_P + \tilde{H}_C + \tilde{H}_I$
into the eigenbasis of $\tilde{H}_0$,
we have $\bar{H} = U_0 \tilde{H} U_0^\dagger $ where, in $\bar{H}_P$
and $\bar{H}_I$ the operator $\sigma_z$ is replaced by $D$.

To remove the time-dependence from our description, we transform the Hamiltonian $\bar{H}$
into a frame rotating at the frequency $\omega_R$ and make a rotating wave approximation.  
Note that in a system with more than two levels,
the choice of a rotating frame is not unique;  here, we choose a rotating frame that allows us to describe
transitions between levels adjacent in energy.  The transition to the rotating frame can be described
using the unitary
\begin{equation}
U_R(t) = \exp\left[-i t \left( \omega_R a^\dagger a +\sum_{n=0}^3 n\omega_R \sigma_{nn} \right)\right].
\end{equation}
In the rotating frame, we have
\begin{equation}
H = U_R \bar{H} U_R^\dagger +i \dot{U}_R U_R^\dagger = H_0 + H_P + H_C + H_I,
\label{HRWA}
\end{equation}
with
\begin{eqnarray}
H_0 &=& \sum_{n=0}^3\left(E_n - n\omega_R\right)\sigma_{nn},\label{H0}\\
H_P &\simeq& \frac{1}{4}\delta\epsilon \sum_{n=0}^2 d_{n+1,n} \sigma_{n+1,n},\label{HP}\\
H_C &=& \Delta_0 a^\dagger a,\label{HC}\\
H_I &\simeq& 2 g_0\left( a\sum_{n=0}^2 d_{n+1,n} \sigma_{n+1,n} + \textrm{h.c.}\right),\label{HI}
\end{eqnarray}
where $\Delta_0=\omega_0 - \omega_R$ is the detuning between the cavity resonance frequency and probe frequency.  Here, $E_n$ and  $d_{n+1,n}=d_{n,n+1}^*$
have to be obtained from the (numerical) diagonalization of $\tilde{H}_0$. The Hamiltoninan described by Eqs.~(\ref{HRWA})--(\ref{HI}) forms the basis
of the following theoretical analysis of the dispersive readout of the valley splitting in a DQD.


\section{Input-output theory}

The response of the DQD system to a microwave probe field can be determined
using input-output theory \cite{Collet1984}.
We begin by finding the stationary solution for the equations of motion of the 
operators $a$ and $\sigma_{n,n+1}$ in the Heisenberg picture, $\dot{a} = i[H,a]$
and $\dot{\sigma}_{n,n+1} = i[H,\sigma_{n,n+1}]$, including the relevant dissipative terms,
\begin{eqnarray}
\dot{a} &=& -i\Delta_0 a -\frac{\kappa}{2}a+\sqrt{\kappa_1} a_{\textrm{in},1}+\sqrt{\kappa_2} a_{\textrm{in},2}\nonumber\\
               & &          -2i g_0\sum_{n=0}^2 d_{n,n+1} \sigma_{n,n+1},\label{eom-a}\\
\dot{\sigma}_{n,n+1}&=& -i (E_{n+1}-E_n-\omega_R)\sigma_{n,n+1} - \frac{\gamma}{2}\sigma_{n,n+1}\nonumber\\
                        & & +\sqrt{\gamma}{\cal F}
                         -2 i g_0 d_{n+1,n}(p_n-p_{n+1}) a,\label{eom-sigma}
\end{eqnarray}
where $\kappa=\kappa_1+\kappa_2+\kappa_i$ is the total cavity decay rate, 
with $\kappa_{1,2}$ the decay rates through the input and output ports, and
$\kappa_i$ the internal decay rate. 
$a_{\textrm{in},1}$ and $a_{\textrm{in},2}$ denote
the incoming parts of the external field at the two ends of the cavity, and
$\gamma$ and ${\cal F}$ are the decay rate and quantum noise
within the DQD. For simplicity we have assumed $\gamma$ to be equal 
for all transitions.
In the following we assume a cavity quality  factor $Q = f_0/\kappa = 2500$
and an electronic dephasing rate of $\gamma = 2.4\,{\rm GHz}$.

A previous work considered the cavity-coupled dynamics with the DQD restricted to the ground state energy level \cite{Petersson2012}. Thermal population of low lying excited states may be important in the Si/SiGe system due to small valley splittings. To account for finite temperature effects, we have replaced the operator $\sigma_{n,n}$ by the occupation probability $p_n = \langle \sigma_{n,n}\rangle$ of the $n^{\rm th}$ DQD level. We assume a thermal population of the DQD levels with 
\begin{equation}
p_n = \frac{e^{-E_n/k_B T}}{\sum_n e^{-E_n/k_B T}}.  \label{Boltzmann}
\end{equation}

The stationary solution is found by setting $\dot{\sigma}_{n,n+1}=0$ in Eq.~(\ref{eom-sigma}),
neglecting the quantum noise ${\cal F}$, and solving for $\sigma_{n,n+1}$, with the result
\begin{equation}
\sigma_{n,n+1} = \frac{-2g_0 d_{n+1,n}(p_n-p_{n+1})}{E_{n+1}-E_n-\omega_R -i\gamma/2} a \equiv \chi_{n,n+1}a ,
\end{equation}
where we have introduced the electric susceptibility $\chi_{n,n+1}$ pertaining to the $n\rightarrow n+1$ transition.
Solving for $a$ in the stationary limit ($\dot{a}=0$) and calculating
the outgoing field $a_{\rm out} = \sqrt{\kappa_2} a$, we find:
\begin{eqnarray}
A = \frac{a_{\rm out}}{a_{\rm in}} &=& \frac{-i\sqrt{\kappa_1 \kappa_2}}
{\Delta_0 - i\kappa/2 +2g_0 \sum_{n=0}^2d_{n,n+1}\chi_{n,n+1}},\quad
\label{output}
\end{eqnarray}
with the real-valued microwave transmission probability 
$|A|^2$ and phase shift $\Delta\phi=-\arg(A)$,
which represents the main analytical result of this paper. In general, the cavity input port is driven with a weak coherent microwave tone, i.e., $a_{\rm in} = \alpha$ with the coherent-state
amplitude $\alpha$.


\section{Results}

Our goal is to extract information about the valley splittings $E_R$ and $E_L$, as well
as the valley-dependent tunneling matrix elements $t$ and $t'$,
from measurements of the cavity transmission. We expect that the electric dipole matrix elements at the avoided crossings 
in Fig.~\ref{fig:cartoon}(b) will lead to features in the amplitude and phase of the 
microwave field transmitted through the DQD. The distances between these four
features are determined by $E_R$ and $E_L$, thus potentially allowing for the extraction of
those two parameters from the analysis of the spectrum. For brevity, we restrict our discussion to the cavity amplitude response. The phase response provides similar information \cite{Petersson2012}.

\begin{figure}
\begin{center}
\includegraphics[width=\columnwidth]{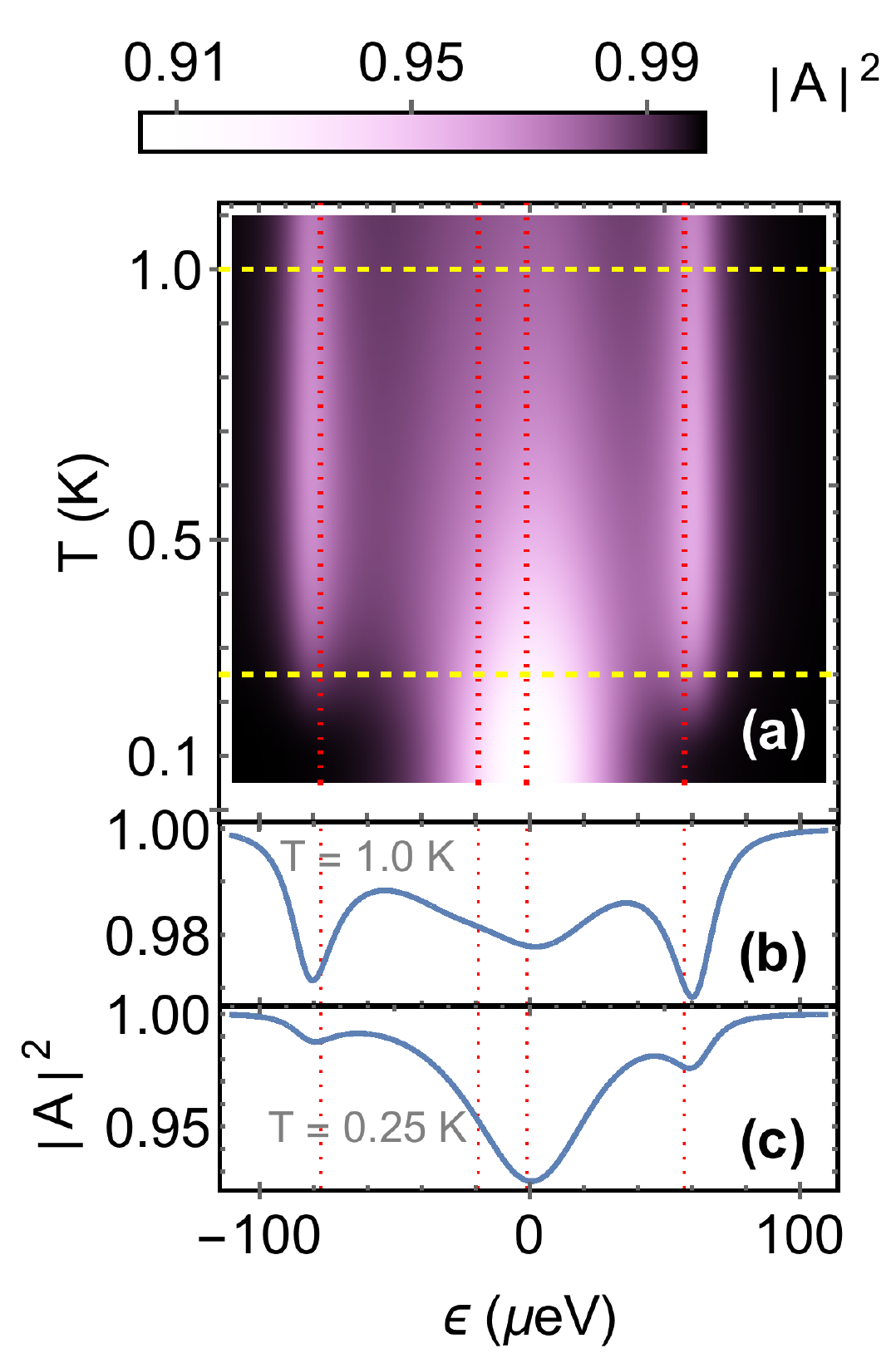}
\caption{(a) Microwave transmission coefficient $|A|^2$ as a function of 
the DQD detuning $\epsilon$ and temperature $T$. The  vertical red dotted lines indicate the 
anticrossings in the DQD spectrum (cf. Fig.~\ref{fig:cartoon}).  The horizontal yellow lines indicate
two cuts at temperatures  $T=1\,{\rm K}$ and $T=250\,{\rm mK}$ for which the transmission
coefficient is plotted separately in (b) and (c).   
The tunneling matrix elements between the QDs are assumed to be 
$t = 25\,\mu {\rm eV}$, while those between opposite valleys are $t' = 13\,\mu {\rm eV}$. The microwave resonator frequency is $f_0 = \omega_0/2\pi = 7.8 \,{\rm GHz} = 32\,\mu{\rm eV}$ and the probe
field is on resonance with the cavity, $\omega_R=\omega_0$.
The valley splittings are chosen as $E_L$ = 76 $\mu$eV, and $E_R$ = 58 $\mu$eV.}

\label{fig:valleystates}
\end{center}
\end{figure}
We first demonstrate that the cavity transmission is sensitive to low lying valley states by evaluating the microwave transmission probability 
$|A|^2$ as a function of $\epsilon$ and $T$ by numerically
diagonalizing $\tilde{H}_0$ for every value of $\epsilon$ and filling the states according
to Eq.~(\ref{Boltzmann}). A two-dimensional plot of $|A(\epsilon,T)|^2$ is 
shown in Fig.~\ref{fig:valleystates}(a). Figures ~\ref{fig:valleystates}(b,c) show cuts through this plot at temperatures of $T=1\,{\rm K}$ and $T=250\,{\rm mK}$.  At low temperatures, most of the population is in the DQD ground-state
and correspondingly, only the lowest avoided crossing near $\epsilon$ = 0 is visible. This avoided crossing results in a reduction in the cavity transmission, as has been observed in GaAs and InAs DQDs \cite{Petersson2012,Frey2012}. As the temperature is increased the population of the higher-lying states increases following Eq.~(\ref{Boltzmann}) and these states start contributing to the cavity response. The $|L'\rangle$-$|R\rangle$ avoided crossing appears as a smaller dip around $\epsilon$ =   -80 $\mu$eV. Due to the smaller left dot valley splitting, the $|L\rangle$-$|R'\rangle$ avoided crossing  has a larger contribution to the cavity response, resulting in a deeper dip in the cavity transmission around $\epsilon$ = 60 $\mu$eV. These simulations demonstrate that valley states can be observed in the cavity response.

\begin{figure}[t]
		\begin{center}
				\includegraphics[width=\columnwidth]{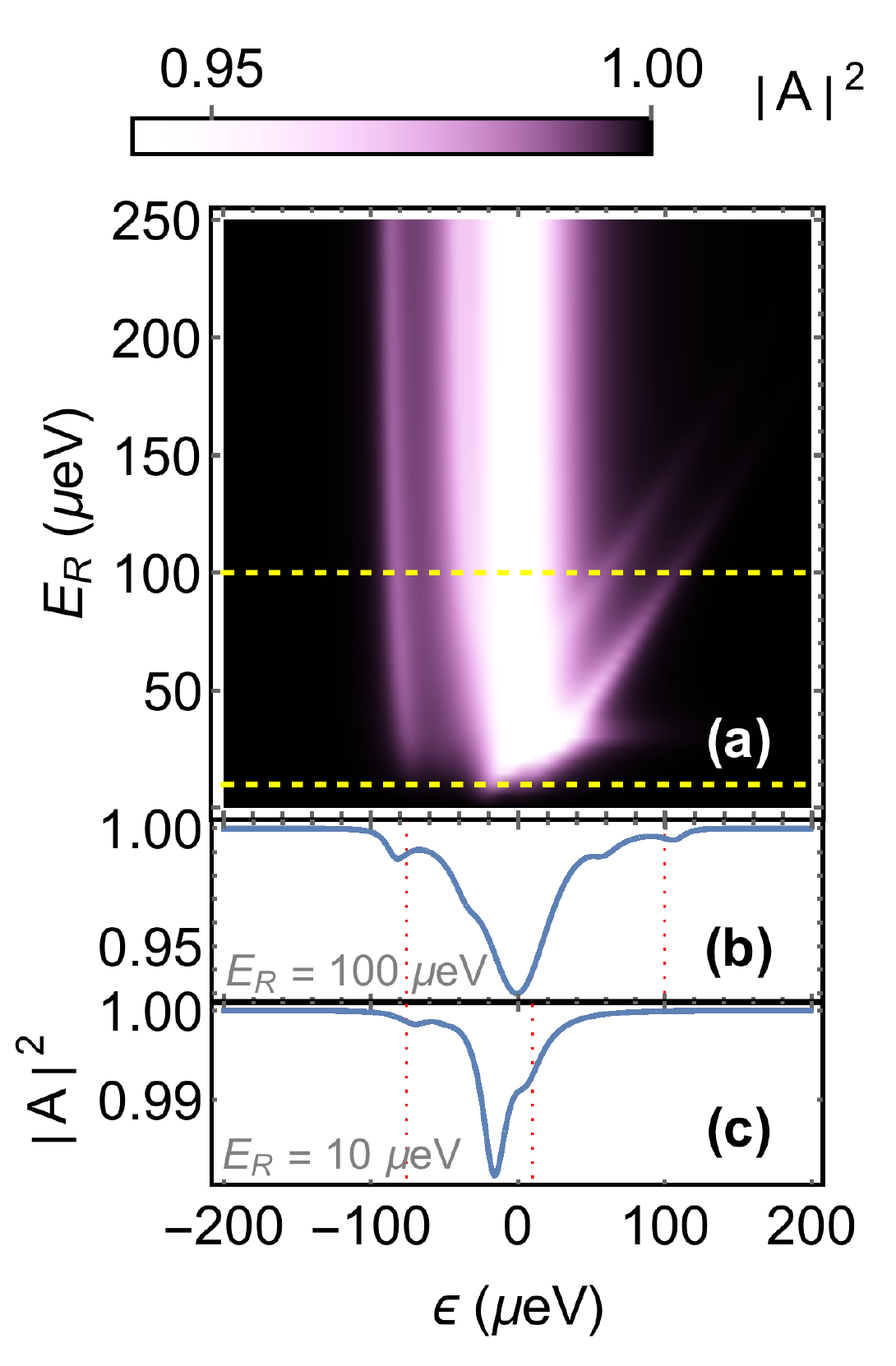}
				\caption{(a) Cavity transmission as a function of the inter-dot bias $\epsilon$ and the right-dot valley splitting $E_R$, with $E_L=76 \,\mu{\rm eV}$, $f_0 = \omega_0/2\pi = 7.8 \,{\rm GHz} = 32\,\mu{\rm eV}$, $\omega_R=\omega_0$, $t$ = 25 $\mu$eV, $t'$ = 13 $\mu$eV, and $T$ = 250 mK. 
The horizontal yellow lines indicate two cuts at valley splittings: (b) $E_R = 100\,\mu{\rm eV}$ and (c) $E_R = 10\,\mu{\rm eV}$.
The red vertical lines indicate $\epsilon=-E_L$ and $\epsilon=E_R$; the other avoided crossings are not shown since they 
are inaccurate due to their vicinity to $\epsilon=0$.}
				\label{fig:valleysplitting}
		\end{center}
\end{figure}

We next show that the cavity response is sensitive to the magnitude of the valley splitting. Figure~\ref{fig:valleysplitting} shows the cavity transmission as a function of detuning and right dot valley splitting. In these simulations the left dot valley splitting is $E_L = 70\, \mu {\rm eV}$ and $T$ = 250 mK.  With $E_R = 0$, the cavity transmission is dominated by the $|L\rangle$-$|R\rangle$ ground state anticrossing and the $|L'\rangle$-$|R\rangle$ anticrossing. Here the cavity response is asymmetric with respect to $\epsilon =0$. As the right dot valley splitting increases, a dip in cavity transmission is observed, which is associated with the $|L\rangle$-$|R'\rangle$ anticrossing. The competition between valley splitting and thermal excitation becomes apparent as the valley splitting is further increased because the cavity response is only sensitive to states that are occupied. As a result, the dip in cavity transmission that is associated with the $|L\rangle$-$|R'\rangle$ anticrossing becomes less pronounced with valley splittings beyond $\approx 150\,\mu{\rm eV}$. A second dip in the cavity response emerges for $\epsilon$ $>$ 0 when $E_R$ $>$ 80 $\mu$eV. This feature is associated with the higher lying $|L'\rangle$-$|R'\rangle$avoided crossing.

\begin{figure}[t]
		\begin{center}
				\includegraphics[width=\columnwidth]{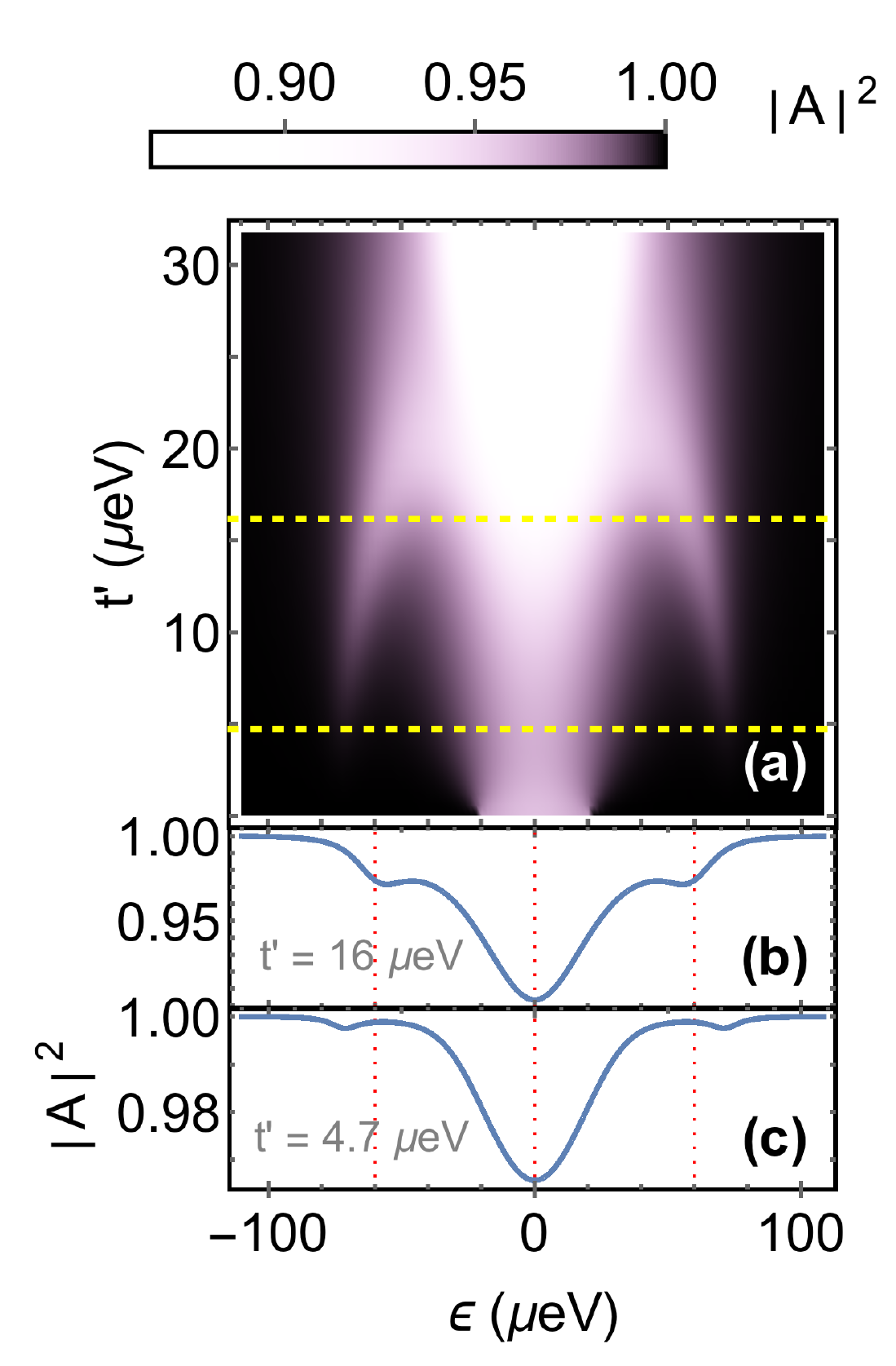}
				\caption{(a) Cavity transmission as a function of $\epsilon$ and the intervalley matrix element  $t'$. The horizontal yellow lines indicate two cuts at intervalley tunnel couplings: (b) $t'=16\,\mu {\rm eV}$ and  (c) $t'=4.7\,\mu {\rm eV}$. Here, we use a fixed $t=25\,\mu {\rm eV}$, $E_L=E_R=60\,\mu {\rm eV}$, and $f_0 = \omega_0/2\pi=7.8 \,{\rm GHz} = 32\,\mu{\rm eV}$, $\omega_R=\omega_0$, $T$ = 250 mK. Red vertical lines as in Fig.~\ref{fig:valleystates}.}
				\label{fig:valleycoupling}
		\end{center}
\end{figure}

In semiconductor DQD charge qubit experiments the interdot tunnel coupling can be tuned using electrostatic gate voltages. Tunability of the tunnel coupling is observed in charge sensing and photon assisted tunneling measurements \cite{DiCarlo2004,Petta2004,Simmons2009}. In contrast, little is known about the experimental tunability of the intervalley coupling. We now show that the cavity response is sensitive to changes in the intervalley matrix element $t'$. The cavity transmission is plotted as a function of $\epsilon$ and $t'$ in Fig.~\ref{fig:valleycoupling}. Here the valley splittings are fixed at $E_R  = E_L = 60\, \mu {\rm eV}$, $t$ = 25 $\mu$eV, and $T$ = 250 mK. For small values of $t'$ the $|L\rangle$-$|R\rangle$ ground state anticrossing dominates the cavity response leading to a significant reduction in the cavity transmission near $\epsilon$ = 0. As $t'$ is increased, the dispersive features associated with the $|L'\rangle$-$|R\rangle$ and $|L\rangle$-$|R'\rangle$ anticrossings broaden and become more pronounced [see Fig.\ ~\ref{fig:valleycoupling}(b)]. The avoided crossings in the energy level diagram begin to merge and are not well defined for $t' > 20\, \mu {\rm eV}$. As a result, a broad dip is observed in the cavity transmission, centered around $\epsilon = 0$. These theoretical predictions show that measurements of the cavity transmission may lead to useful characterization of the intervalley coupling rate.


\section{Conclusions}
cQED-based approaches to quantum information science have been very productive \cite{DevoretReview2013}. They have allowed long-range coupling of qubits, high fidelity readout of cavity coupled quantum devices, and investigations of mesoscopic physics. In this paper, we have demonstrated that the cQED architecture can be used as a sensitive probe of low-lying valley states. For realistic device parameters, the cavity transmission exhibits dips that are associated with energy level anticrossings with low lying valley states. The position of these dips in cavity transmission yields the valley splittings. The temperature dependence of the cavity response also gives information on the magnitude of the valley splitting. Since the cavity probes the susceptibility of the DQD, it is also sensitive to the curvature of the energy levels, and can be used to extract the intervalley matrix elements. Due to the high energy resolution of narrow-band microwave spectroscopy, coupling Si/SiGe QDs to microwave cavities may allow for efficient measurements of valley splittings in an approach that is complementary to existing approaches, such as magnetospectroscopy \cite{Borselli2011,Yang2012} or the quantum Hall effect \cite{Goswami2007,Mi2015}.  The method presented in this paper can potentially be applied to probe the energy level structure of different types of quantum dots, e.g. Zeeman energies for spin sublevels in a gradient field \cite{Pioro2008} or spatially varying g-factors in strong spin-orbit systems \cite{Nadj2012}.


\begin{acknowledgements}
Funded by the Army Research Office through grant No.\ W911NF-15-1-0149 with partial support from the Gordon and Betty Moore Foundation's EPiQS Initiative through Grant GBMF4535, the National Science Foundation (DMR-1409556 and DMR-1420541), and the German Research Foundation (SFB 767).
\end{acknowledgements}


\appendix


\section{Valley-dependent model of a double quantum dot}
\label{DQDmodel}

Here, we derive our model Eq.~(\ref{Hdqd}) for a single electron in a DQD
with a single valley-degenerate orbital in each QD.
We start from a description of the DQD in a common valley basis for both QDs
because it allows us to formulate a model for valley-preserving tunneling 
through the smooth electrostatic barrier between the two QDs.
We then obtain Eq.\ (\ref{Hdqd}) by changing into the local valley eigenbasis in each QD.

The state of the electron on the left (right) QD is denoted $|l\rangle$ ($|r\rangle$), and
we introduce the Pauli operators in this left-right orbital Hilbert space as
$\sigma_z |l\rangle = +|l\rangle$ and $\sigma_z |r\rangle = -|r\rangle$,
and $\sigma_x |l\rangle = |r\rangle$, etc.
The valley-independent part of the DQD Hamiltonian can be written as
\begin{equation}
H_d = \frac{\epsilon}{2} \sigma_z + t_c \sigma_x,
\end{equation}
where $\epsilon$ represents the DQD energy detuning (bias) energy and
$t_c$ the inter-dot tunneling matrix element which we can choose to be real.

The two low-energy valley states in the extended
two-dimensional electron system in a Si/SiGe quantum well
are denoted $|\pm z\rangle$, and we introduce the corresponding
valley Pauli operators $\tau_z'|\pm z\rangle = \pm|\pm z\rangle$,
$\tau_x'|+z\rangle = |-z\rangle$, etc.
The most general two-level valley Hamiltonian for each
of the two individual QDs can then be written as
 $ {\bm \delta}_i\cdot {\bm \tau}'/2$ where $i=l,r$
and ${\bm \delta}_i$ is an arbitrary vector whose length
determines the bare valley splitting in QD $i$.
Using the sum and difference 
${\bm \delta}_\pm = ( {\bm \delta}_l \pm {\bm \delta}_r$)/2, 
we can write the valley Hamiltonian of the DQD as
\begin{equation}
H_v' = \frac{1}{2}\sum_{i=l,r} |i\rangle\langle i| \, {\bm
  \delta}_i\cdot {\bm \tau}'
= \frac{1}{2} \left({\bm \delta}_+ +  \sigma_z  {\bm \delta}_- \right) \cdot {\bm \tau}' .
\end{equation}
Rotating the valley basis such that the common $\tau_z$ valley quantization axis is parallel to ${\bm \delta}_+$
and the $\tau_x$ axis along the projection of ${\bm \delta}_-$ into the plane perpendicular to  ${\bm \delta}_+$,
we find  
\begin{equation}
H_v' = \frac{\delta}{2} \tau_z + \frac{1}{2}\left(\delta_z \tau_z +\delta_x \tau_x \right)\sigma_z.
\end{equation}
Combining $H_d$ and $H_v'$, we obtain
\begin{equation}
H_0'  =  H_d+ H_v' =\begin{pmatrix}
H_L & t_c \openone \\
t_c \openone & H_R
\end{pmatrix},
\label{Hdqd1}
\end{equation}
where $\openone$ denotes the 2x2 identity matrix and
\begin{eqnarray}
H_L &=& \frac{\epsilon}{2}+\frac{E_L}{2} +\frac{E_L}{2} \begin{pmatrix}
\cos\theta_L  & \sin\theta_L  \\
\sin\theta_L  & -\cos\theta_L 
\end{pmatrix}, \\              
H_R &=& -\frac{\epsilon}{2}+\frac{E_R}{2} +\frac{E_R}{2} \begin{pmatrix}
\cos\theta_R  & \sin\theta_R  \\
\sin\theta_R  & -\cos\theta_R 
\end{pmatrix}, 
\end{eqnarray}
with the valley splittings and angles
\begin{eqnarray}
\Delta_L &=& \delta+\delta_z,   \label{DeltaL}\\
\Delta_R &=& \delta-\delta_z,\label{DeltaR}\\
E_{L,R} &=& \sqrt{\Delta_{L,R}^2+\delta_x^2},\label{ELR}\\
\tan\theta_{L,R} &=& \frac{\delta_x}{\Delta_{L,R}}.\label{thetaLR}
\end{eqnarray}
We have shifted the definition of $\epsilon$ by $(E_L-E_R)/2$ and omitted an irrelevant constant energy shift by $\sqrt{\delta^2 + \delta_x^2}$, in order to 
center the level crossing of the lower valley eigenstates in the left and right QD at $\epsilon=0$ and zero energy.
In order to obtain Eq.~(\ref{Hdqd}), we rotate the valley basis about the y-axis by $\theta_i$ in QD $i$, using 
the transformation
\begin{equation}
U = \begin{pmatrix}
\cos \frac{\theta_L}{2}  & \sin \frac{\theta_L}{2}  & 0 & 0\\
-\sin \frac{\theta_L}{2} & \cos \frac{\theta_L}{2} & 0 & 0\\
0 & 0 & \cos \frac{\theta_R}{2}  & \sin \frac{\theta_R}{2}  \\
0 & 0 & -\sin \frac{\theta_R}{2} & \cos \frac{\theta_R}{2} 
\end{pmatrix},
\end{equation}
and obtain
\begin{equation}
\tilde{H}_0 = U H_0' U^\dagger = \begin{pmatrix}
\epsilon/2+E_L & 0 & t & t' \\
0 & \epsilon/2        &  -t' & t \\
 t   &      -t'            & -\epsilon/2+E_R & 0 \\
  t'  &      t             & 0 &  -\epsilon/2  
\end{pmatrix},
\label{HdqdA}
\end{equation}
with
\begin{eqnarray}
t = t_c \cos\left(\frac{\theta_L + \theta_R}{2}\right),\label{thorizontal}\\
t' = t_c \sin\left(\frac{\theta_L + \theta_R}{2}\right). \label{tcross}
\end{eqnarray}
Here, since $t_c$ was chosen real, both $t$ and $t'$ will be real.
The Hamiltonian Eq.~(\ref{HdqdA}) is expressed in the local valley eigenbasis,
\begin{eqnarray}
|L'\rangle &=& |l\rangle \left( \cos \frac{\theta_L}{2}|+z\rangle  + \sin \frac{\theta_L}{2}|-z\rangle \right),\\
|L\rangle &=& |l\rangle \left ( -\sin \frac{\theta_L}{2}|+z\rangle  + \cos \frac{\theta_L}{2}|-z\rangle \right),\\
|R'\rangle &=& |r\rangle \left ( \cos \frac{\theta_R}{2}|+z\rangle  + \sin \frac{\theta_R}{2}|-z\rangle \right),\\
|R\rangle &=& |r\rangle \left ( -\sin \frac{\theta_R}{2}|+z\rangle  + \cos \frac{\theta_R}{2}|-z\rangle \right).
\end{eqnarray}

\bibliography{Burkard_valley_v7}

\end{document}